\RequirePackage{fix-cm}

\documentclass{svjour3} 
\usepackage{verbatim}
\usepackage{cite}
\usepackage{graphicx}
\usepackage{graphics}
\usepackage{amssymb}
\usepackage{color}
\usepackage{doi}
\usepackage[margin=2cm]{geometry}

\smartqed  

\begin{document}

\title{BCS model on Quasiperiodic Lattices}

\author{T. F. A. Alves  \and
        F. W. S. Lima   \and
        A. Macedo-Filho \and
        G. A. Alves     }

\institute{T. F. A. Alves \at
          Departamento de F\'{\i}sica, Universidade Federal do Piau\'{\i}, 57072-970, Teresina - PI, Brazil \\
          \email{tay@ufpi.edu.br} 
   \and
          F. W. S. Lima \at
          Departamento de F\'{\i}sica, Universidade Federal do Piau\'{\i}, 57072-970, Teresina - PI, Brazil
   \and
          A. Macedo-Filho \at
          Campus Prof.\ Antonio Geovanne Alves de Sousa, Universidade Estadual do Piau\'{\i}, 64260-000, Piripiri - PI, Brazil
   \and
          G. A. Alves \at
          Departamento de F\'{\i}sica, Universidade Estadual do Piau\'{\i}, 64002-150, Teresina - PI, Brazil
}

\date{Received: date / Accepted: date}

\maketitle

\begin{abstract}

We study the Biswas-Chatterjee-Sen (BCS) model, also known as the KCOD (Kinetic Continuous Opinion Dynamics) model on quasiperiodic lattices by using Kinetic Monte Carlo simulations and Finite Size Scaling technique. Our results are consistent with a continuous phase transition, controlled by an external noise. We obtained the order parameter $M$, defined as the averaged opinion, the fourth-order Binder cumulant $U$, and susceptibility $\chi$ as functions of the noise parameter. We estimated the critical noises for Penrose, and Ammann-Beenker lattices. We also considered 7-fold and 9-fold quasiperiodic lattices and estimated the respective critical noises as well. Irrespective of rotational and translational long-range order of the lattice, the system falls in the same universality class of the two-dimensional Ising model. Quasiperiodic order is irrelevant and it does not change any critical exponents for BCS model.


\end{abstract}

\section{Introduction}

Over the years, Physics has been developing its potential in the interdisciplinary field, with contributions in several areas, such as biology, chemistry, engineering, sociology, etc. In sociophysics\cite{Galam2008-1, Galam2008-2,Stauffer2008,SenCambridge}, which is an interdisciplinary field between physics and sociology, physics seeks to provide a qualitative analysis of phenomena arising from the emergent properties of a group of individuals/agents that interact with each other.

Sociophysics is recent and still in a stage of development\cite{Stauffer2008,SenCambridge}. Anyways, its main objective is to seek general standards that characterize social relations. These general patterns can be obtained from the analysis of the phase transitions and critical behavior presented by some consensus formation models\cite{Brady2003, Lima2012, Pereira2005, Stanley2018, Vieira2016, Vilela2009, Wu2009}. In this line, in 2012, Biswas, Chatterjee and Sen\cite{BISWAS20123257} proposed a consensus formation model, namely the BCS model, to study opinion dynamics by pairwise interactions, and inspired by wealth exchange models\cite{ChatterjeeEPJB,ChakrabartiCambridge}.

In the definition of the BCS model, we can define a ``society'', with its individuals are placed each one on a lattice vertex, having one continuous opinion state ranging from the extremal opinion $-1$ to $+1$, representing an one-dimensional political opinion spectra regarding a particular subject. Because of the model comprises continuous opinion states, it is also known as the KCOD (kinetic continuous opinion dynamics) model\cite{lima2017-1, lima2017-2, Anteneodo2014, Anteneodo2017}. Opinion states are initially selected at random and then, we let the system evolve to an stationary state according to its kinetic rules. The opinions of the selected individuals are updated by means of an interaction term or affinity with one randomly selected nearest neighbor, that can be negative with a probability controlled by a noise parameter. According to BCS model, consensus in a society can be reached by means of a system with $N$ individuals/agents interacting between closest neighbor pairs, when the noise is below a critical threshold.

Another example of consensus formation model is the Majority vote model\cite{MJOliveira1992}, where the randomly chosen individual interacts with all of his closest neighbors. In both models, we can define a noise parameter, modelling local discordances. In the Archimedean lattices\cite{Lima2012, Yu2017, BISWAS20123257, PhysRevE.94.062317}, BCS and Majority Vote models present continuous phase transitions with a well defined critical noise. Both models are in the same universality class of the two-dimensional (2D) Ising model. Both models are outside from equilibrium, being defined by an master equation rathen than a Hamiltonian, turning both models interesting in the point of view of non-equilibrium models and critical phenomena\cite{PhysRevE.94.062317}.

However, some questions remain open about these non-equilibrium models. Among them, we consider if the quasiperiodic order changes the critical exponents. Harris-Barghathi-Vojta criterion\cite{Vojta2009, Barghathi2014}, answers the question, by stating that the connectivity disorder is irrelevant if $a\nu > 1$, where $\nu=1$ for BCS model\cite{BISWAS20123257, PhysRevE.94.062317} in the square lattice and $a$ is the disorder decay exponent, defined by the scaling relation $z = L^{-a}$, with $\Delta z$ being the standard deviation of the mean vertex coordination number $z$. The exponent $a$ is $3/2$ for quasiperiodic lattices and consequently, they have irrelevant disorder according to Harris-Barghathi-Vojta criterion. Therefore, our main objective is to investigate if the criterion predicts the correct critical behavior for BCS model. In addition, we can expect that the quasiperiodic order is irrelevant because there is no example of a model that changes its critical behavior when coupled with quasiperiodic lattices, compared to the periodic ones.

In the next section, we define the BCS model, and then, in section 3, we show our numerical results and discussions. We conclude by present our conclusions in section 4.

\section{Model and Simulations}

We consider the BCS model, coupled to Penrose, Ammann-Beenker, 7-fold, and 9-fold lattices. The kinetic Monte Carlo rules of BCS model are written as follows\cite{BISWAS20123257}:
\begin{enumerate}
\item For each vertex $i$ of the lattice with $N$ vertexes, is assigned one individual with a continuous opinion variable $o_{i}(t)$, in the interval $[-1,1]$. We start the dynamics by selecting the opinion state for each vertex of the lattice using $N$ random numbers generated in the interval $[-1,1]$;
\item For each time step, we randomly select a lattice vertex to be updated;
\item Next, we randomly select only one of its bonds and define the affinity $\mu_{i,j}$ of the bond. The affinity parameter is a continuous random variable in the interval [0,1]. The affinity can be turned negative with a probability $q$. The parameter $q$ acts as an external noise, modelling local discordances;
\item The two vertexes $i$ and $j$ that share the selected bond are updated according to the following
\begin{eqnarray}
o_{i}(t+1) &=& o_{i}(t) + \mu_{i,j}o_{j}(t), \nonumber \\
o_{j}(t+1) &=& o_{j}(t) + \mu_{i,j}o_{i}(t), 
\label{opinion_t+1}
\end{eqnarray}
where the variables $o_{i}(t)$ and $o_{j}(t)$ are the previous opinion states while the $o_{i}(t+1)$ and $o_{j}(t+1)$ stand for the updated opinion states of the two vertexes $i$ and $j$, respectively. 
\item If any updated opinion state is higher (lower) than $+1$ ($-1$), then it is made equal to $+1$ ($-1$), to preserve the limit $[-1,1]$. Note that this introduces non linearity to the model.
\end{enumerate}

\begin{figure}[h]
\begin{center}
\includegraphics[scale=0.15]{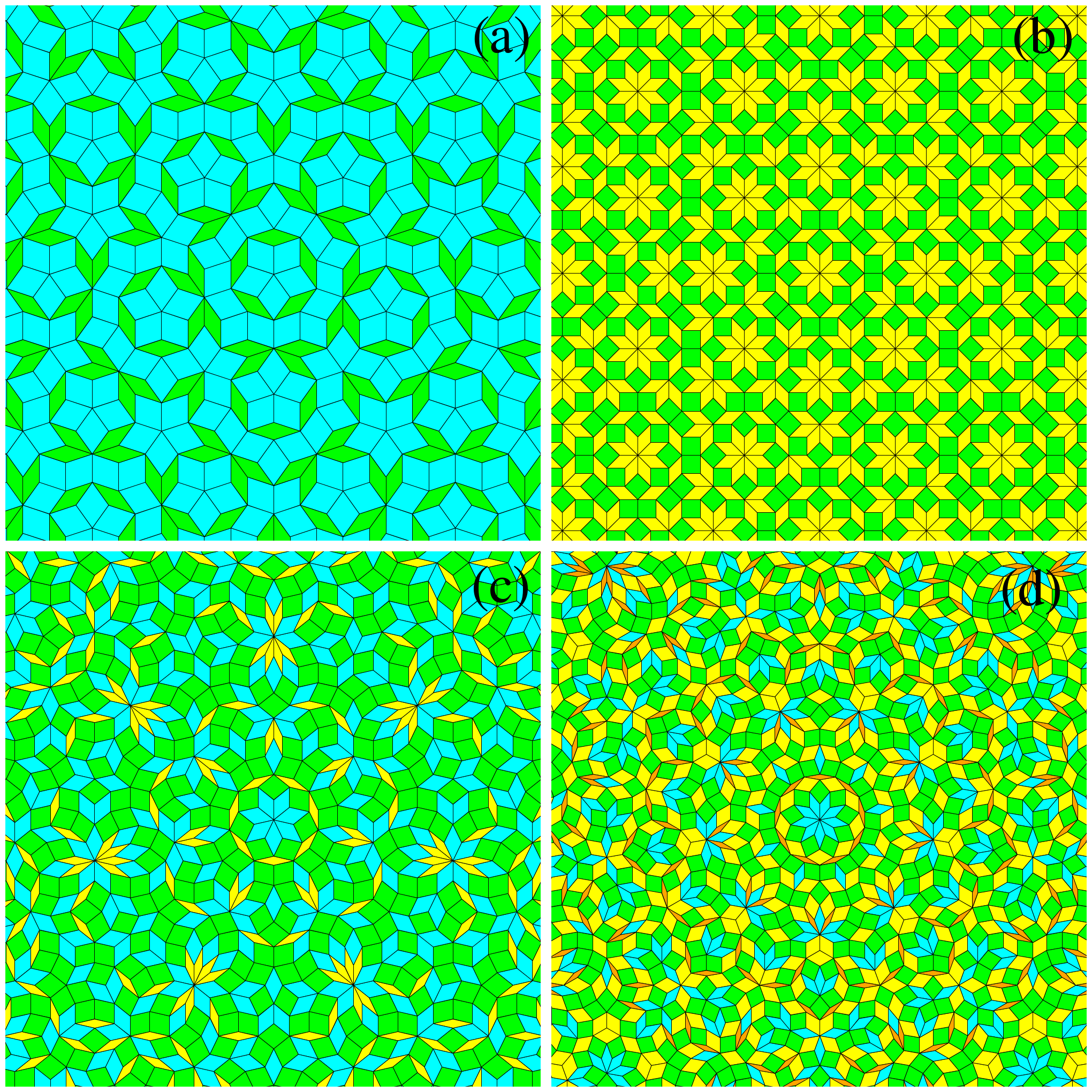} 
\end{center}
\caption{(Color Online) (a) 5-fold Penrose tiling. (b) 8-fold Ammann-Beenker tiling. (c) 7-fold tiling. (d) 9-fold tiling.}
\label{quasilattices}
\end{figure}

We show in Fig.(\ref{quasilattices}), sections of the four quasiperiodic lattices used in our BCS model simulations. We used the Projection method\cite{Naumis2003}, with some modifications to include the bonds, in order to generate quasiperiodic lattices\cite{Mota2018}. This method is justified by de Bruijn theorem\cite{deBruijn1981}, which states that any quasiperiodic lattice generated by a pentagrid projection is the Penrose tiling. From generalization of pentagrids to multigrids, it is possible to generate $n$-fold tilings. The number of lattice vertexes $N$ of tilings is a function of the lattice generation $g$. The lattice generation $g$ is related to the number of parallel lines of the multigrid\cite{Mota2018}.

Now we turn our attention to the necessary observables to describe the BCS model critical behavior. Consensus is measured by the following parameter
\begin{equation}
O = \frac{1}{N} \sum_i^N o_i
\label{consensus-balance}
\end{equation}
where $N$ is the total number of lattice vertexes. If we increase the noise parameter $q$, the model can present a symmetry breaking transition, that is, below a particular value $q_{c}$ of the $q$ parameter, the system is ordered (giving a nonzero, finite value of the order parameter $O$), while a disordered phase exists above $q_{c}$ ($O=0$). According to the previous discussion\cite{BISWAS20123257, PhysRevE.94.062317}, we can observe that the BCS model presents a phase transition very similar to the ferromagnetic-paramagnetic phase transition in magnetic systems.

The order parameter $M(q)$, susceptibility $\chi(q)$, and Binder's fourth-order cumulant $U(q)$ are given by the following relations, respectively \cite{Lima2012}
\begin{eqnarray}
M(q)= \langle O \rangle, \label{orderparameter}\\
\chi(q) = N (\langle O^{2} \rangle - \langle O \rangle ^{2}), \label{orderparameterfluctuations} \\
U(q) = 1 - \frac{\langle O^{4} \rangle}{3\langle O^{2} \rangle^{2}}, \label{4-order-cumulant}
\end{eqnarray}
where $O$ is given on Eq.(\ref{consensus-balance}) and the brackets $\langle ... \rangle$ represents the average of a time series, obtained by a Monte Carlo Markov Chain (MCMC). Quantities described by Eqs. (\ref{orderparameter}), (\ref{orderparameterfluctuations}), and (\ref{4-order-cumulant}) are functions of the noise parameter $q$ and, close to the critical noise, they should obey the following finite-size scaling relations\cite{Lima2012}
\begin{eqnarray}
M = N^{-\beta/2\nu}f_{M}\left(N^{1/2\nu}(q-q_{c})\right), \label{orderparameter-fss}\\
\chi = N^{\gamma/2\nu}f_{\chi}\left(N^{1/2\nu}(q-q_{c})\right), \label{orderparameterfluctuations-fss} \\
U = f_{U}\left(N^{1/2\nu}(q-q_{c})\right),\label{4-order-cumulant-fss}
\end{eqnarray}
respectively, where $\nu$, $\beta$, and $\gamma$ are the critical exponents and, for 2D systems like quasiperiodic tilings, the characteristic length of the system is $L=N^{1/2}$.
 
To obtain the quantities above, we performed MCMC's on lattices shown in Fig.(\ref{quasilattices}). Lattice sizes (defined as the number of vertexes $N$) used in the majority of our numerical results are given in Tab.(\ref{alllatticesizes}), as functions of the lattice generation parameter $g$. In our simulations, we used $2 \times 10^{5}$ MCMC steps to allow the system reaching the stationary state, and then $1 \times 10^{7}$ steps at the stationary state to collect the time series to calculate all needed averages. Statistical errors were obtained by using ``jackknife'' resampling technique\cite{Tukey1958}.

\begin{table}[h]
\caption{Number of vertexes $N$ of quasiperiodic lattices as function of lattice generation parameter $g$.}
\label{alllatticesizes} 
\begin{tabular}{ccccc}
\hline\noalign{\smallskip}
Generation         & Penrose         & Ammann-Beenker  &  7-fold        & 9-fold            \\
\noalign{\smallskip}\hline\noalign{\smallskip}
$g=5$              & $N=871$         & $N=2121$        & $N=1751$       & $N=2854$          \\
$g=6$              & $N=1221$        & $N=2929$        & $N=2339$       & $N=3862$          \\
$g=7$              & $N=1596$        & $N=3801$        & $N=3081$       & $N=5095$          \\
$g=8$              & $N=2001$        & $N=4873$        & $N=3949$       & $N=6490$          \\
$g=9$              & $N=2476$        & $N=6065$        & $N=4866$       & $N=8011$          \\
$g=10$             & $N=3041$        & $N=7409$        & $N=5902$       & $N=9703$          \\
\noalign{\smallskip}\hline
\end{tabular}
\end{table}

\section{Results and Discussion}

In this section, we show our numerical results and determine the system universality class. First, we will discuss the difficulty that consensus formation models have when it comes to estimating critical exponents by using finite-size scaling regressions, directly related to the precision in determining the critical point. The results for Majority Vote model, presented by reference\cite{Yu2017} show that if the critical point is estimated with a relative error of $0.0005$ or greater, the estimates for the critical exponents $\beta/\nu$ and $\gamma/\nu$ deviate $8\%$ and $2\%$ or worse, respectively. In order to solve this difficulty, the reference\cite{Yu2017} used a very high statistical precision that required $10^{9}$ terms of the Markovian time series.
        
Following the discussion, we suggest another method that does not require a simulation with this number of terms of the time series\cite{Alves_2019}. We suggest measuring the critical threshold by comparing the data collapses obtained by changing the critical noise values, close to the average Binder cumulant crossing value. We then search for the critical noise value that determines the best collapse with the 2D Ising model exponents. The best data collapse is one that goes on for a longer interval, with noise values farther away from the possible critical point. With this suggestion we can reduce computational time. If collapses with the 2D Ising model fail, a better statistics with a longer Markovian time series is then required, as done by reference\cite{Yu2017} to rule a new universality class. Our present numerical results allowed us to estimate the critical noises with 4 significant figures, which means that the uncertainties are equal to $0.0005$ in all cases.

\begin{figure}[p]
\begin{center}
\includegraphics[scale=0.2]{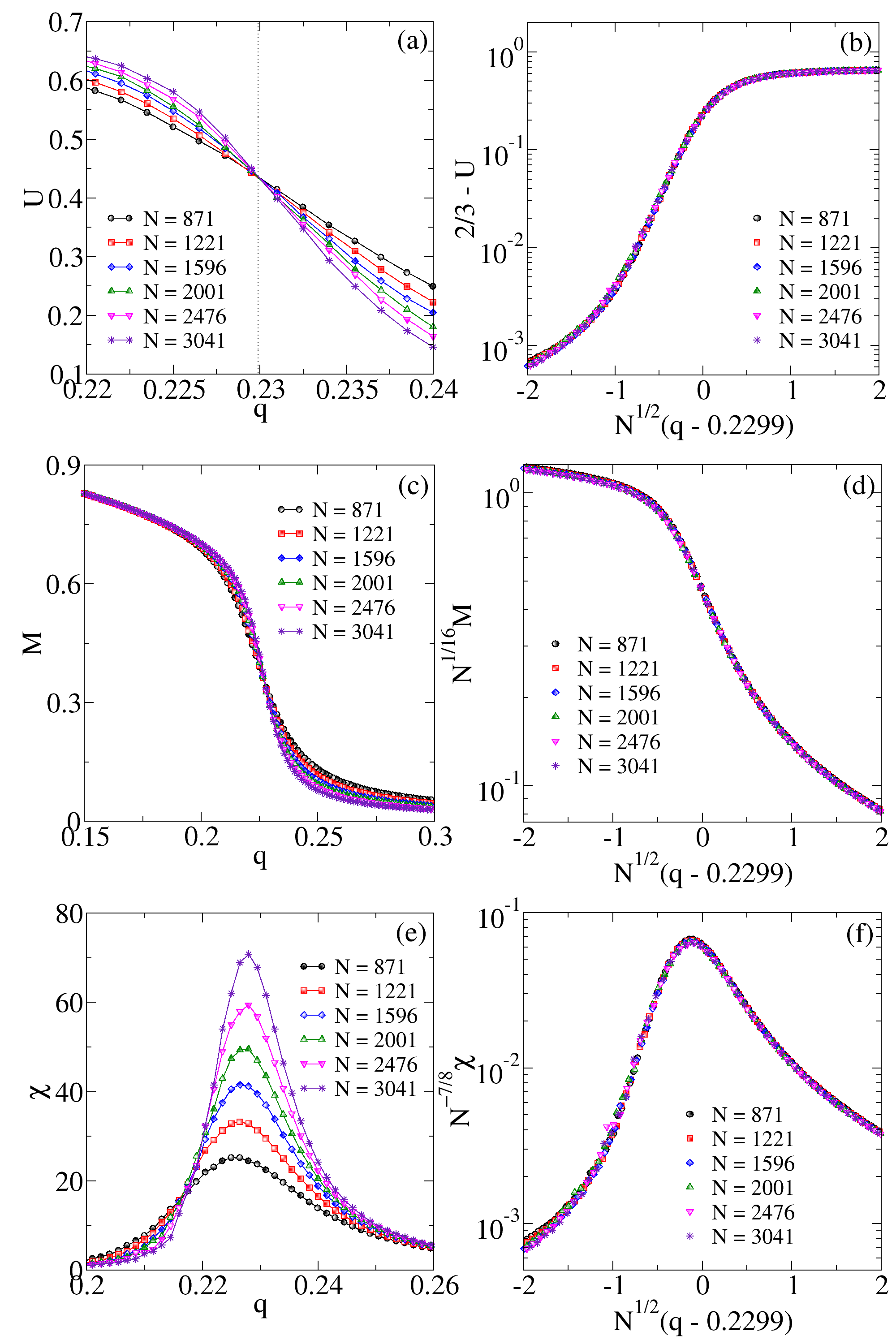}
\end{center}
\caption{(Color Online) Averages for different sizes $N$ of Penrose lattice, showed on Table (\ref{alllatticesizes}) and respective data collapses by using the 2D Ising critical exponent ratios $1/\nu=1$, $\beta/\nu=1/8$ and $\gamma/\nu=7/4$. Note the logarithm scale on the abscissa of rescaled plots. (a) Binder cumulant $U$ as function of noise $q$. The curve crossings of the Binder cumulant allow for estimating the critical noise value. According to the data collapses, our most accurate estimate of the critical noise is $q_{c} \approx 0.2299$. (b) Binder cumulant $U$ versus scaling variable $N^{1/2\nu}(q-q_{c})$. (c) Order parameter $M$ as function of noise $q$. Order parameter curves suggest a continuous phase transition. (d) Order parameter rescaled by $N^{\beta/2\nu}$ versus scaling variable $N^{1/2\nu}(q-q_{c})$. (e) Susceptibility $\chi$ as function of noise parameter $q$. Susceptibility maxima increase with $N$, meaning a divergence at the critical threshold in the infinite lattice size limit. (f) Susceptibility rescaled by $N^{-\gamma/2\nu}$ versus scaling variable $N^{1/2\nu}(q-q_{c})$. Statistical error bars are smaller than symbols.}
\label{penrose_fig}
\end{figure}

\begin{figure}[p]
\begin{center}
\includegraphics[scale=0.2]{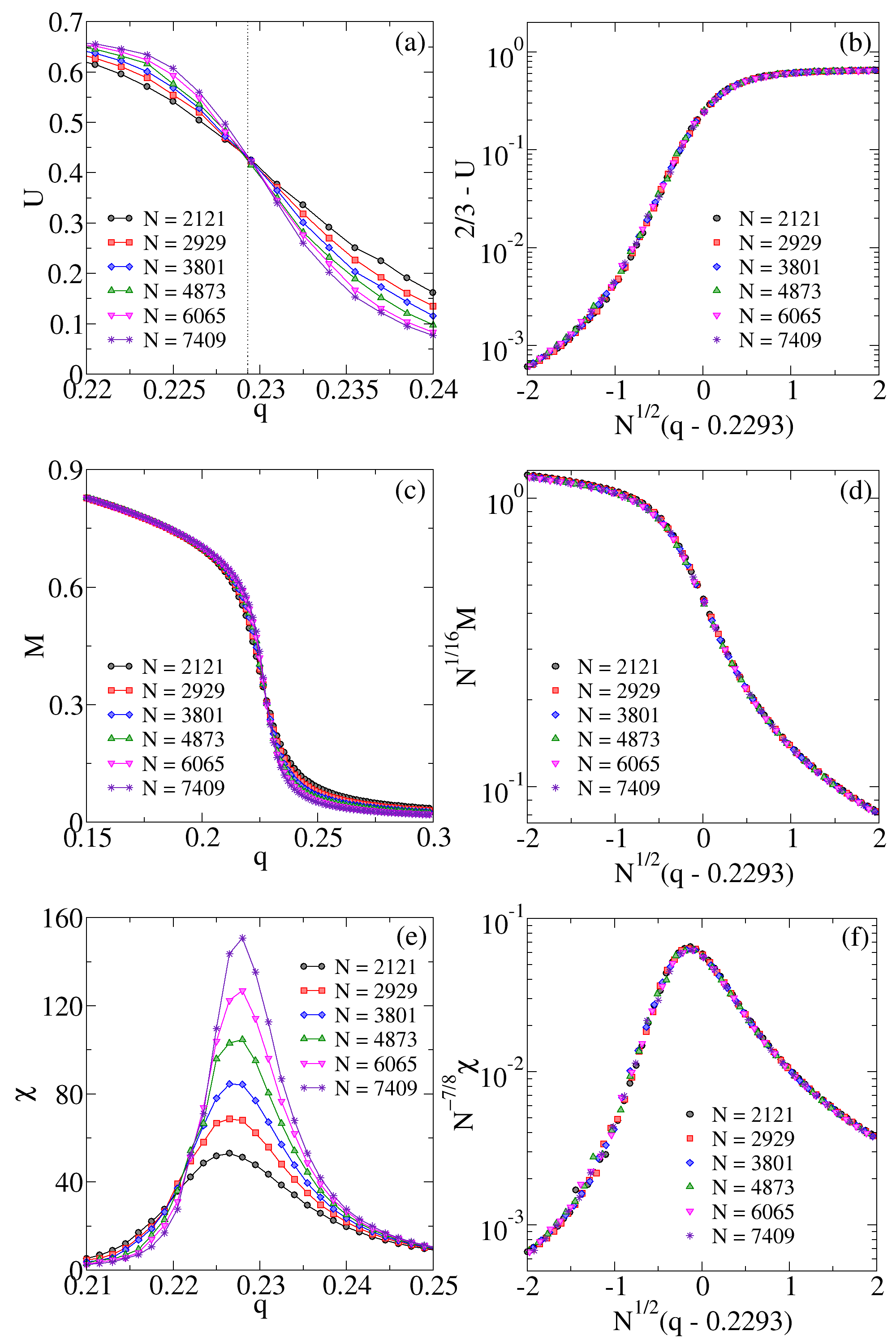}
\end{center}
\caption{(Color Online) Similar to Fig.(\ref{penrose_fig}) but for Ammann-Beenker lattice. Our estimative for the critical noise is $q_{c} \approx 0.2293$.}
\label{ammannbeenker_fig}
\end{figure}

\begin{figure}[p]
\begin{center}
\includegraphics[scale=0.2]{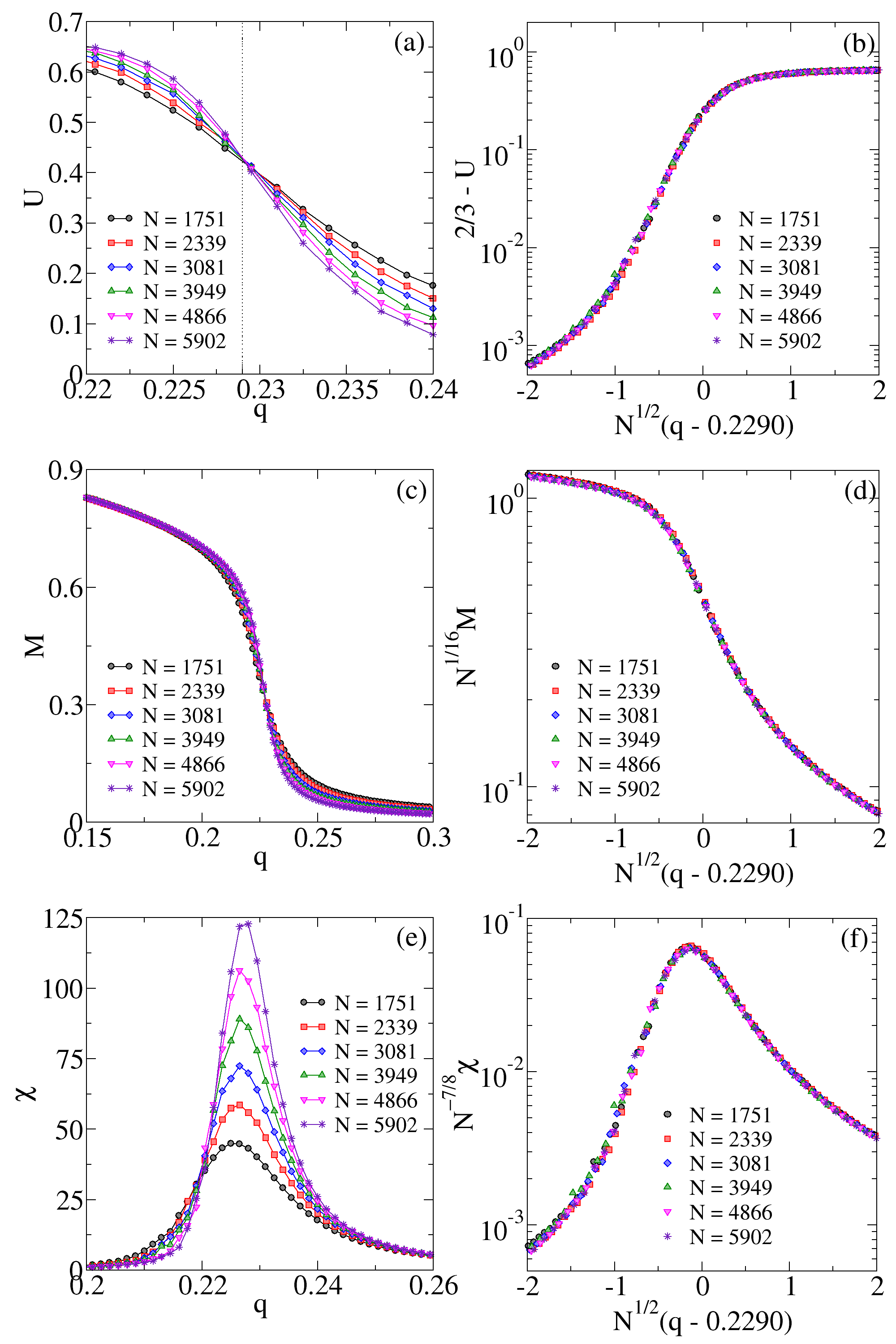}
\end{center}
\caption{(Color Online) Similar to Fig.(\ref{penrose_fig}) but for 7-fold lattice. Our estimative for the critical noise is $q_{c} \approx 0.2290$.}
\label{7fold_fig}
\end{figure}

\begin{figure}[p]
\begin{center}
\includegraphics[scale=0.2]{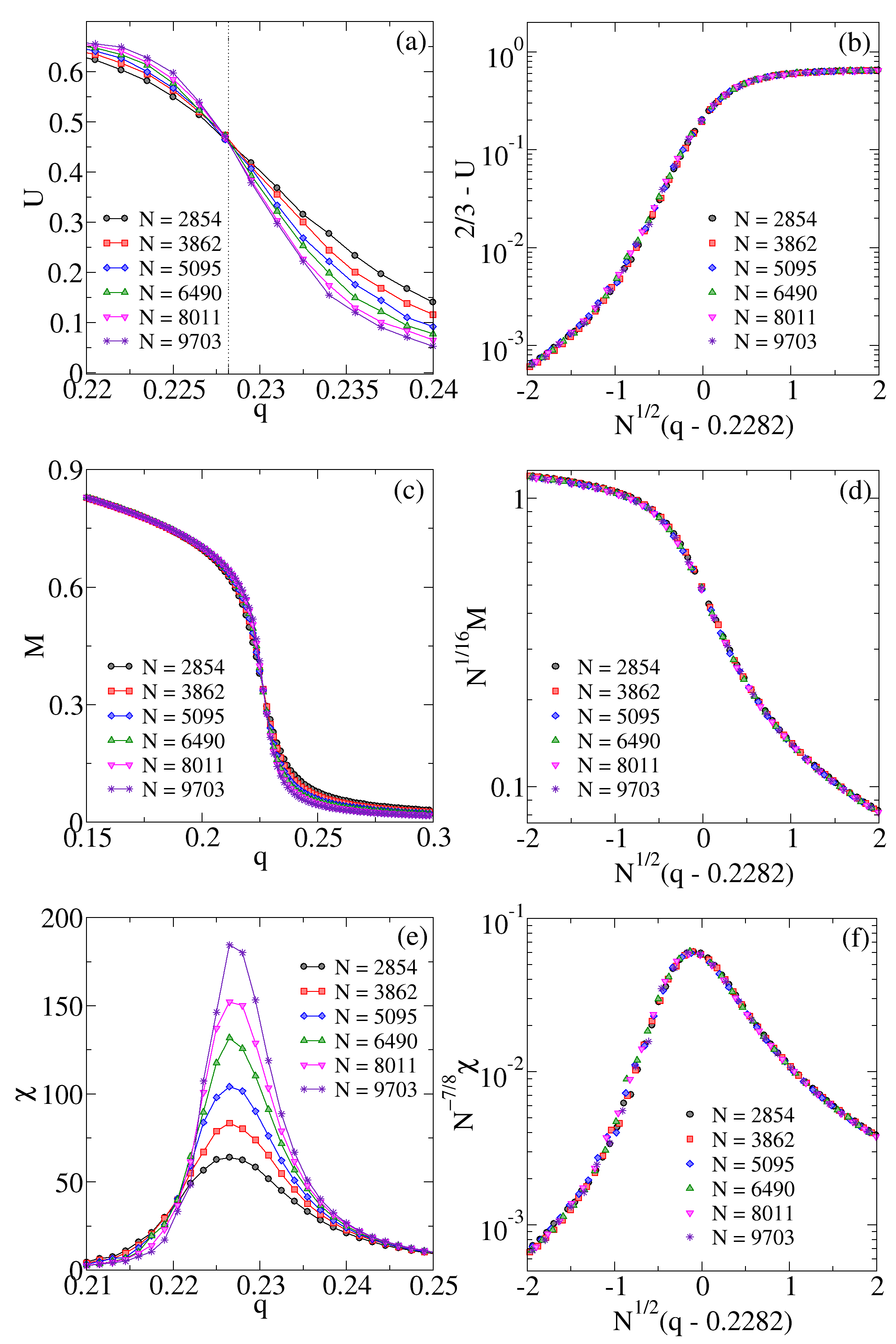}
\end{center}
\caption{(Color Online) Similar to Fig.(\ref{penrose_fig}) but for 9-fold lattice. Our estimative for the critical noise is $q_{c} \approx 0.2282$.}
\label{9fold_fig}
\end{figure}

We first considered Binder cumulant $U$, presented in Figs. (\ref{penrose_fig}-a), (\ref{ammannbeenker_fig}-a), (\ref{7fold_fig}-a), and (\ref{9fold_fig}-a) as functions of noise parameter $q$ for Penrose, Ammann-Beenker, 7-fold and 9-fold lattices, respectively, for each lattice generation given in Tab.(\ref{alllatticesizes}). The estimated values for critical noises are shown on Tab.(\ref{allcriticalnoises}). The estimates are all close to the average crossing value, and were obtained by the analysis of the best data collapses by rescaling the axes according to scaling relations written on Eq.(\ref{4-order-cumulant-fss}) together with 2D Ising critical exponent ratios $1/\nu=1$, $\beta/\nu=1/8$ and $\gamma/\nu=7/4$. Binder cumulant collapses are shown in Figs. (\ref{penrose_fig}-b), (\ref{ammannbeenker_fig}-b), (\ref{7fold_fig}-b), and (\ref{9fold_fig}-b) for Penrose, Ammann-Beenker, 7-fold, and 9-fold lattices, respectively.

\begin{table}[h]
\caption{Estimated critical noises.}
\label{allcriticalnoises} 
\begin{tabular}{cc}
\hline\noalign{\smallskip}
Quasiperiodic Lattices & Aproximated Critical Noises \\
\noalign{\smallskip}\hline\noalign{\smallskip}
Penrose                & $q_{c} \approx 0.2299$      \\
Ammann-Beenker         & $q_{c} \approx 0.2293$      \\
7-fold                 & $q_{c} \approx 0.2290$      \\
9-fold                 & $q_{c} \approx 0.2282$      \\
\noalign{\smallskip}\hline
\end{tabular}
\end{table}

Next, we show in Figs.(\ref{penrose_fig}-c), (\ref{ammannbeenker_fig}-c), (\ref{7fold_fig}-c), and (\ref{9fold_fig}-c), our results for the order parameter as function of the noise parameter $q$ for Penrose, Ammann-Beenker, 7-fold and 9-fold lattices, respectively. We can observe that the curves of the order parameter are compatible with a continuous phase transition from the ordered phase for values less than $q_{c}$ to a disordered phase, for values greater than $q_{c}$. Figs. (\ref{penrose_fig}-d), (\ref{ammannbeenker_fig}-d), (\ref{7fold_fig}-d), and (\ref{9fold_fig}-d) show respective data collapses of the order parameter curves by using the scaling relations written in Eq.(\ref{orderparameter-fss}) for Penrose, Ammann-Beenker, 7-fold, and 9-fold lattices, respectively. Again, the best data collapses were obtained by using the critical noises on Tab.(\ref{allcriticalnoises}) and the critical exponents of the 2D Ising model.

Finally, we plot the susceptibility $\chi$ as a function of the noise parameter $q$, as shown in the Figs. (\ref{penrose_fig}-e), (\ref{ammannbeenker_fig}-e), (\ref{7fold_fig}-e), and (\ref{9fold_fig}-e) for Penrose, Ammann-Beenker, 7-fold, and 9-fold lattices, respectively. In lattices with infinite size limits, we can observe that the susceptibility $\chi$ diverge close to their respective critical thresholds $q_{c}$, suggesting also a continuous phase transition. For finite lattices sizes, the maximum values of susceptibility move towards the critical thresholds and increase with the lattice size $N$. In Figs. (\ref{penrose_fig}-f), (\ref{ammannbeenker_fig}-f), (\ref{7fold_fig}-f), and (\ref{9fold_fig}-f) we show our best data collapses of the susceptibility curves obtained by using the critical noise estimates on Tab.(\ref{allcriticalnoises}) and the critical exponents of the 2D Ising model.

In fact, our results suggest that the system falls on the 2D Ising universality class when coupled with quasiperiodic lattices, in the same way of Majority Vote model\cite{Alves_2019}, placing the two models on the same universality class as seen for Archimedean lattices\cite{Lima2012, Yu2017, BISWAS20123257, PhysRevE.94.062317}. In general, we expect that the BCS and Majority Vote models should fall in the same universality class, even in more complex topologies. Its particularly interesting that BCS model falls into 2D Ising universality class, even for a continuous opinion variable, which is an evidence of the Ising universality class robustness.

\section{Conclusions}

We investigated, by using Kinetic Monte Carlo simulations, the BCS model coupled to quasiperiodic lattices. According to Harris-Barghathi-Vojta criterion \cite{Vojta2009,Barghathi2014}, the quasiperiodic order is irrelevant and the critical exponents should not change in presence of a quasiperiodic long range order. We numerically obtained the Binder cumulant, the order parameter, and its susceptibility. We estimated the critical noises for Penrose, Ammann-Beenker, 7-fold, and 9-fold lattices at $q_{c} \approx 0.2299$, $q_{c} \approx 0.2293$, $q_{c} \approx 0.2290$, and $q_{c} \approx 0.2282$ respectively. Our numerical results suggest that the system falls on 2D Ising model universality class, despite the quasiperiodic order, in agreement with Harris-Barghathi-Vojta criterion.

\section{Acknowledgments}

We would like to thank CNPq (Conselho Nacional de Desenvolvimento Cient\'{\i}fico e tecnol\'{o}gico), CAPES (Coordenação de Aperfei\c{c}oamento de Pessoal de N\'{\i}vel Superior), FUNCAP (Funda\c{c}\~{a}o Cearense de Apoio ao Desenvolvimento Cient\'{\i}fico e Tecnol\'{o}gico) and FAPEPI (Funda\c{c}\~{a}o de Amparo a Pesquisa do Estado do Piau\'{\i}) for the financial support. We acknowledge the use of Dietrich Stauffer Computational Physics Lab, Teresina, Brazil, and Laborat\'{o}rio de F\'{\i}sica Te\'{o}rica e Modelagem Computacional - LFTMC, Piripiri, Brazil, where the numerical simulations were performed.

\section*{Conflict of interest}

The authors declare that they have no conflict of interest.

\bibliographystyle{./spphys.bst}
\bibliography{textv1}

\end{document}